\documentclass{cernyrep} 
\usepackage{texnames}
\usepackage[T1]{fontenc}
\begin{document}
\title{Toward an {\em Ab-initio} Description of Quasiparticle Properties
    \footnote{Proceedings of the  {\em 12th International Conference
     on Nuclear Reaction Mechanisms}, Varenna, June 15-19, 2009.}}
 
\author{C. Barbieri}

\institute{Theoretical Nuclear Physics Laboratory, RIKEN Nishina Center, 2-1 Hirosawa, Wako, Saitama 351-0198 Japan}

\maketitle 

\begin{abstract}
 Preliminary \emph{ab-initio} applications of many-body Green's functions theory to the ground state of $^4$He suggest that high accuracy can be achieved in the so-called Faddeev-random-phase-approximation method.
 We stress the potentialities of this approach for microscopic studies of medium-large nuclei and report applications to $1s0d$ and $1p0f$-shell nuclei. In particular, we discuss the role of long-range correlations on spectroscopic factors and their dependence on asymmetry.
\end{abstract}
 
\section{Introduction}

The advent of radioactive ion beams has opened the possibility of reaching unexplored regions
of the nuclear landscape. This has lead to the realization that single particle levels evolve with changing
the number of neutrons or protons~\cite{sch.02,taka.05}, sometimes invalidating the conventional magic numbers
known for stable isotopes.
 Another unexpected observation is that spectroscopic factors for the same states (which we will
also refer to as \emph{quasiparticles}) may change dramatically with proton/neutron asymmetry~\cite{gad.08}.
This implies non trivial (and not yet well understood) changes in nuclear correlations at the driplines.
This talk reports about on going investigations of the evolution of quasiparticle properties
using large scale microscopic calculations.

From the point of view of \emph{ab-initio} nuclear structure, great progresses have been made
for light nuclei. There exist now a wide range of techniques that solve the few-body  problem exactly~\cite{he_bench}.
 Moving to medium and large nuclei is made difficult by the exponential grow in degrees of freedom. In this case, techniques like coupled-cluster (CC)~\cite{hag.07,hag.08,hag.09} and self-consistent Green's functions (SCGF)~\cite{dicvan,bar.06,bar.09c} are preferable. For both methods, the number of configurations to be dealt with scales more gently with the increasing particle numbers and the infinite summation of connected diagrams guarantees size extensivity and accuracy.
Applications to medium-heavy nuclei are (for the moment) limited to systems around closed subshell structures.

For open shell systems, standard shell model calculations remain the best option. However, one faces questions about what effective interactions and charges have to be used when extrapolating to the driplines.
One could attempt to bridge shell model Hamiltonians to realistic nuclear forces by employing the above \emph{ab-initio} approaches and studying quasiparticle states and their interactions for different shell closures across the nuclear chart.
 In this respect, the SCGF is the method of choice since the one-body Green's function already contains, in its energy representation, all information of quasiparticle properties. As we will note below,  self-consistency (SC)  helps in improving accuracy and implies the fulfillment of conservation laws.  Moreover, the strong connection between the Green's functions formalism and the experimental data allows to gain unmatched insight into the nuclear dynamics.

\begin{figure}[t]
  \includegraphics[height=.21\textheight]{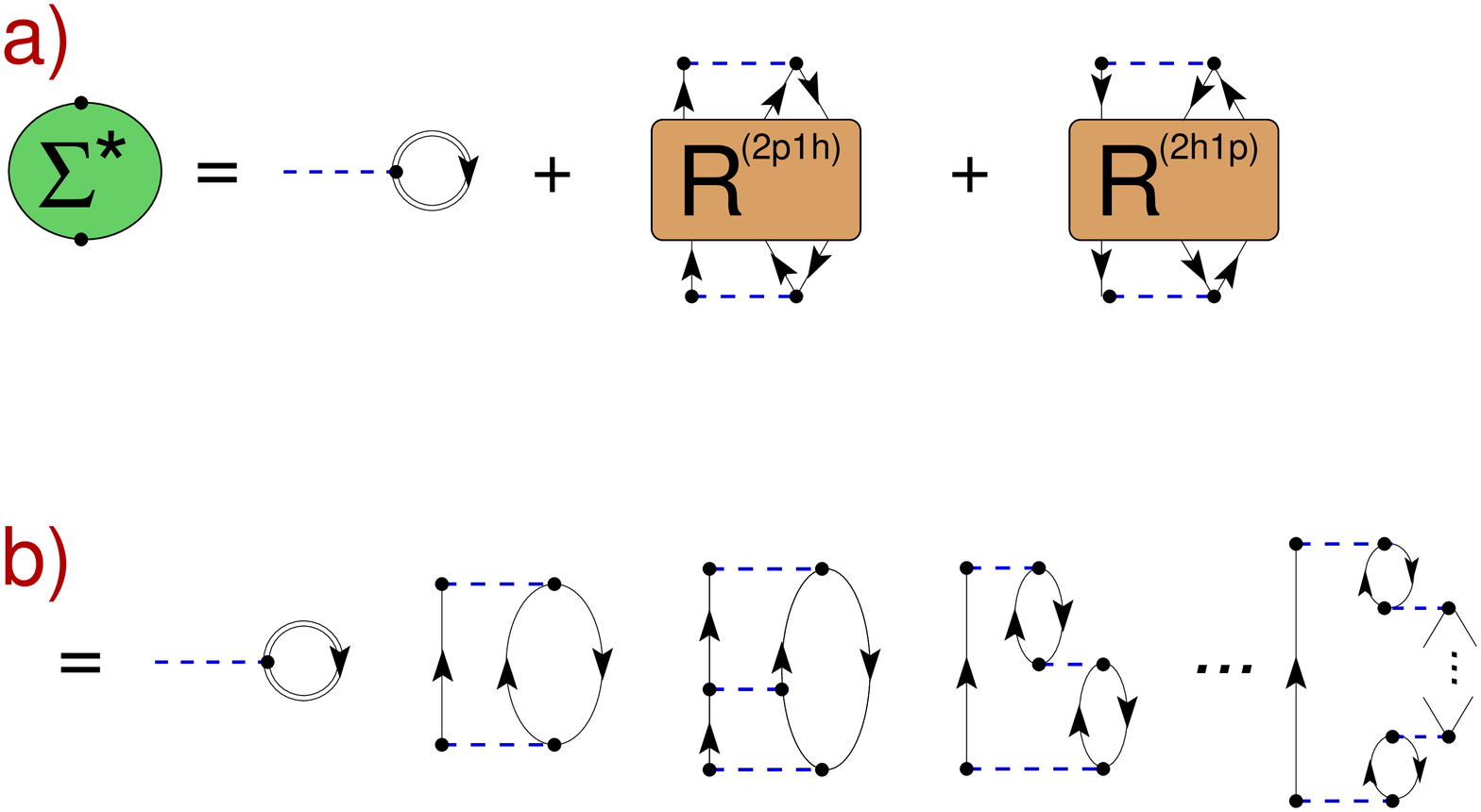}
  \hspace{1.5cm}
  \includegraphics[height=.19\textheight]{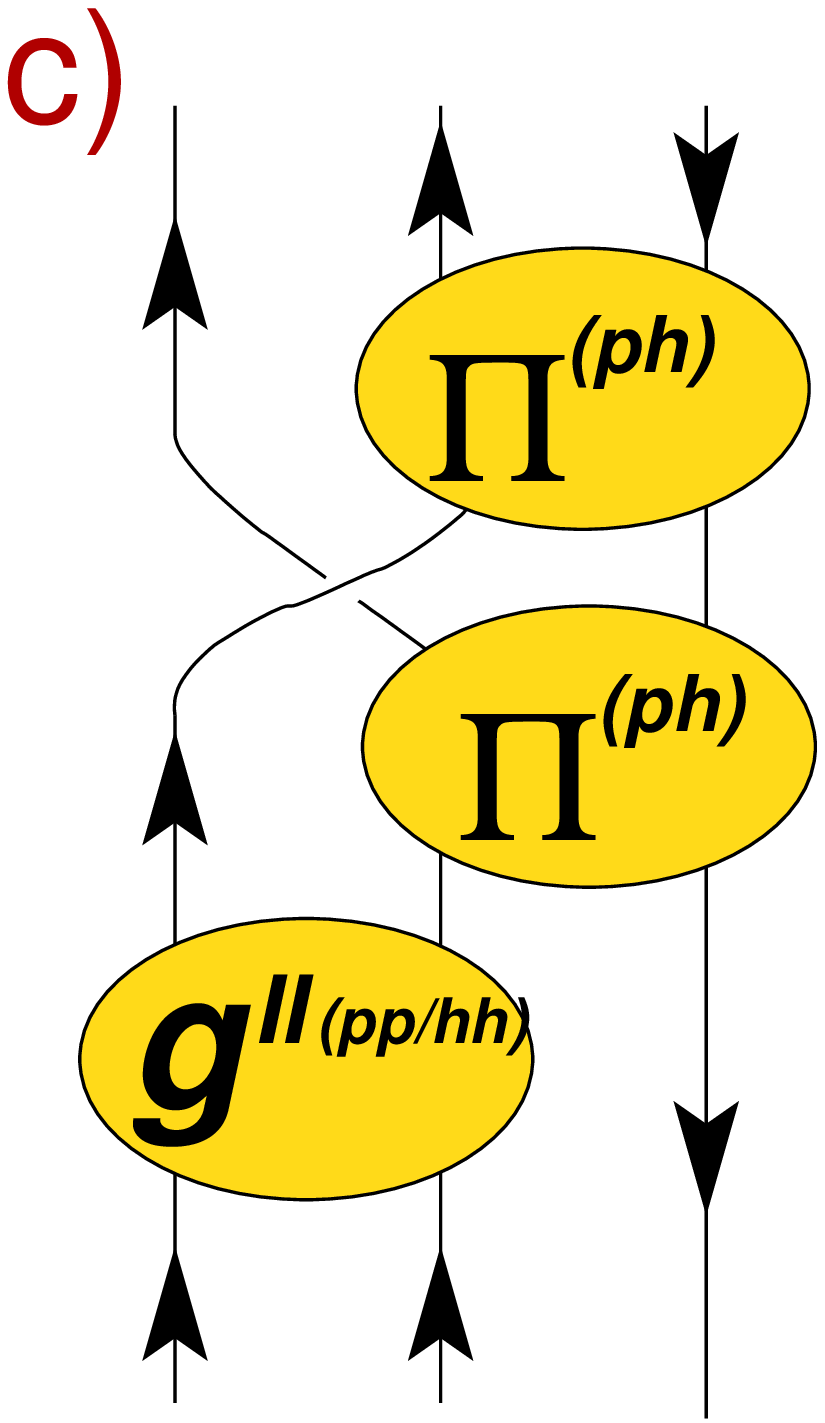}
  \caption{The self-energy $\Sigma^\star(\omega)$ separates exactly into
    a mean field term and the polarization propagators $R(\omega)$ for the
    2p1h/2h1p motion, as shown in a). The double lines represent fully correlated
    propagators. Upon expansion of $R(\omega)$ in Feynman diagrams, one obtains the 
    series of diagrams b) for the self-energy.
     The diagram c) gives an example of the contributions to $R^{2p1h}(\omega)$
    that are summed to all orders by the FRPA method.
    \label{frpa_diags}
     }
\end{figure}

The state of the art SCGF calculations employ the so-called Faddeev random phase approximation (FRPA) to expand the nuclear self-energy. This method is detailed in~\cite{bar.01,bar.07,bar.09c} and is summarized diagrammatically in Fig.~\ref{frpa_diags}. The $\Pi(\omega)$ and $g^{II}(\omega)$ propagators carry information on the collective motion of particle-hole (giant resonance) and two-particle or two-hole (pairing like) configurations.
The interaction between nucleons and collective excitations of the nucleus~\cite{bar.01,bar.07} are accounted for by resumming infinite series of Feynman diagrams, like the one depicted in Fig.~\ref{frpa_diags}c.
The inclusion of RPA phonons makes this method particularly suited for medium-heavy isotopes, where collective states are important. Nevertheless, we will show in Sec.~\ref{sec:he4} that rather accurate results are possible even for nuclei as small as $^4$He. We will further discuss the accuracy and convergence of quasiparticle energies in Sec.~\ref{sec:spe}. And finally review ongoing applications to spectroscopic factors in $1s0d$ and $1p0f$ shell systems, including cases of large proton-neutron asymmetry.

\section{\emph{Ab-initio} Calculations of $^4$He}
\label{sec:he4}

The spectral function of $^4$He has been computed in an harmonic oscillator basis using the $V_{lowk}$ interaction~\cite{Bog.03,Bog.07} (derived from the Argonne {\em v18} potential with a sharp cutoff of $\Lambda$=~1.9~fm$^{-1}$).
We solved for the FRPA self-energy including all contributions to third oder, as explained in Ref.~\cite{bar.07}, and used the Koltun sum rule to extract the binding energy~\cite{bof.71,dicvan}. The calculations employed the intrinsic Hamiltonian, $H_{int}=T+V-T_{cm}$, to remove the contribution of the kinetic energy of the center of mass.

\begin{figure}[h]
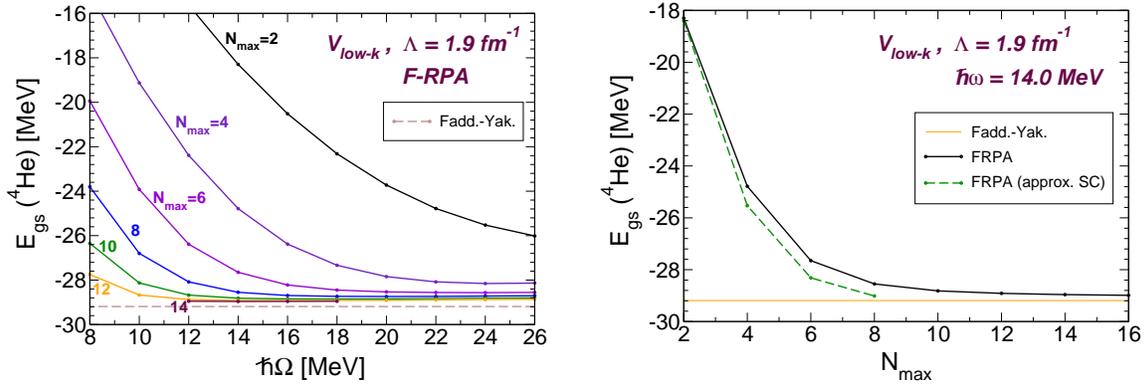

\begin{center}
\includegraphics[width=0.45\columnwidth,clip=true]{He4_vs_hw_Nmax__VlowkL19.eps}
  \hspace{.5cm}
\includegraphics[width=0.45\columnwidth,clip=true]{He4_vs_Nmax__hw14__VlowkL19.eps}
\end{center}
\caption{Convergence of the FRPA binding energy for $^4$He.
  Left: dependence on the oscillator parameter and basis size (without self-consistency). Right:
  convergence for fixed $\hbar\Omega$=14~MeV (black line) and corresponding
  correction due to approximate self-consistency (dashed green line). The horizontal lines mark the
  exact Faddeev-Yakubowsky result.
  }
\label{fig:he4}
\end{figure}

Figure~\ref{fig:he4} shows the convergence of the total energy with respect to the oscillator frequency $\hbar\Omega$ and the basis size. Up to 17 shells ($N_{max}$=16) and partial waves with $l$$\leq$~5 were used. For the largest space, results are basically converged (to within 10 keV) and approach exponentially to -29.00~MeV.  This has to be compared with the exact Faddeev-Yakubowsky result of~-29.19(5)~keV~\cite{hag.07}. An estimate of the self-consistency effects---given by the dashed line in the right panel---further lowers the total energy and brings it closer to the exact result.
At the time of writing, the converged result for full self-consistency is not known and more work is in progress to answer this question~\cite{bar.09.he4}.

We note that, based on the perturbative analysis of Ref.~\cite{tro.05}, the third order algebraic diagrammatic construction method [ADC-(3)], the similar \emph{non}-self-consistent FRPA, and coupled-cluster theory with singles and doubles (CCSD) are all expected to give similar results for the binding energy.
Doing self-consistency in Green's functions theory introduces implicitly several other diagrams. Among these are the same  fourth and fifth order diagrams evaluated by the triple corrections to the CC method [CCSD(T)].
This is indeed the picture that can be inferred from comparing Fig.~\ref{fig:he4} and the analogous CC study of Ref.~\cite{hag.07}. Further confirmation comes from analogous studies of atomic systems~\cite{bar.09a}.

Self-consistency introduces effects of correlations that are beyond the bare FRPA level and are expected to improve sensibly the accuracy of results. Following this approach has two main advantages. First, the corrections are included in the self-energy and therefore apply to all the quantities derived from it. These include the total energy, quasiparticle energies, overlap wave functions, scattering observables and so on. Second, it is known from the theorems of Baym and Kadanoff~\cite{bay.61,bay.62} that this procedure selects those additional diagrams that allow to satisfy general conservation laws.


\section{Calculations of Quasiparticle States}
\label{sec:spe}

\begin{figure}[b]
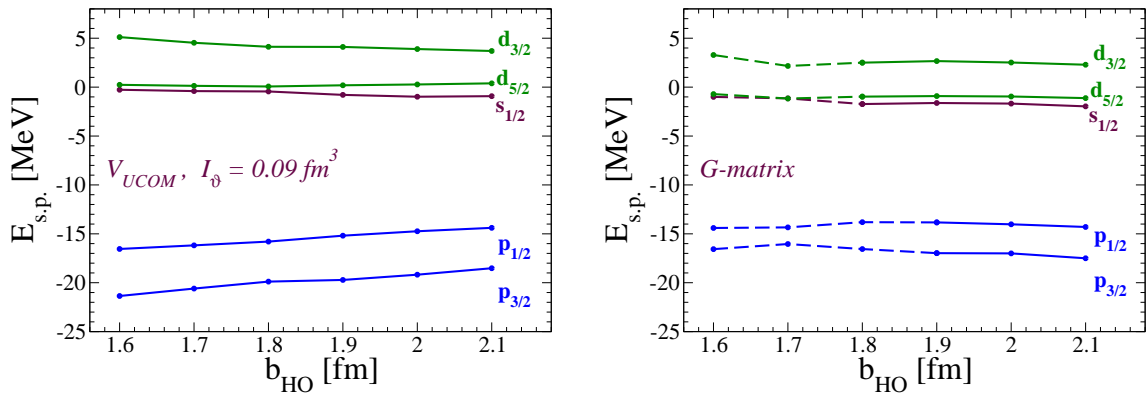

\begin{center}
\includegraphics[width=0.45\columnwidth,clip=true]{Levels_Vvsb.eps}
  \hspace{.5cm}
\includegraphics[width=0.45\columnwidth,clip=true]{Levels_Gvsb.eps}
\end{center}
\caption{Neutron quasiparticle energies for $^{16}$O, as obtained
for the $V_{\text{UCOM}}$~(left) and the Argonne {\em v18}~(right) interactions~\cite{bar.06}.
 For the latter, a G-matrix resummation of short-range diagrams was employed. 
  }
\label{fig:O16spe}
\end{figure}

The self-consistent FRPA approach has been employed in Ref.~\cite{bar.06} to study the single particle spectrum around the Fermi surface of $^{16}$O. A large oscillator basis of eight shells was employed. The resulting energies for the addition or removal of a nucleon are shown in Fig.~\ref{fig:O16spe} as a function of the oscillator length.  The two panels show the results obtained for the regularized $V_{\text{UCOM}}$~\cite{rot.05} interaction and the Argonne {\em v18}~\cite{wir.95} potential (both without the Coulomb force). For the sole case of the Argonne potential, the G-matrix technique was employed to resum ladder diagrams from outside the model space. This approach allows to properly include the effects of short-range correlations~\cite{bar.09c,geu.96}.
 While quasiparticle and quasihole energies obtained with $V_{\text{UCOM}}$ still carry some dependence on the oscillator parameter, the corresponding spin-orbit slitting appears to be rather well converged. For the Argonne {\em v18} calculation, the dependence on the oscillator parameter is sensibly weaker, due to the inclusion of ladder diagrams from outside the model space.

The results for spin-orbit splittings of the $0p$ and $0d$ orbits are reported in Tab.~\ref{sosplit} and compared to the experimental values for neutron single particle states (as observed in $^{15}$O and $^{17}$O). The UCOM renormalization procedure tames the short-range correlations at the level of two body forces and enhances the non-locality of the interaction.
This leads to larger spin-orbit splittings than the {\em v18} case and closer to the experiment.
However, the energy gaps between quasiparticle energies and the $0p_{1/2}$ hole are overestimated by the $V_{\text{UCOM}}$ interaction. This is a consequence of the well know issue of nuclear radii, which are usually underestimated by this interaction without three-nucleon forces~\cite{rot.06}.
The total energy of $^{16}$O with this interaction has also been studied more recently, using similar model spaces~\cite{rot.09}.
 Note that the $0p$ splitting of 3.1~MeV obtained for the Argonne {\em v18} potential is very close to the variational Monte Carlo result of 3.4~MeV (which was derived for the older {\em v14} model)~\cite{pie.93}. This gives further confirmation of the accuracy of our calculations. The remaining discrepancy with respect to the experiment should not be attributed to the many-body method but to the neglect of three-nucleon forces in the Hamiltonian.
The {\em v18} interaction with a G-matrix treatment of the short-range physics reproduces particle-hole (ph) gaps in much better agreement with the observation.

\begin{table}[b]
\begin{center}
\begin{tabular}{llcccccc}
\hline
\hline
     &                           &~&   $V_{\text{UCOM}}$  &~&   Argonne {\em v18} &~&  \emph{ Exp.} \\
     &                           &~&          &~&   (G-matrix)  &~&              \\
\hline
s.-o. splittings: &
    $\Delta E_{p_{1/2}-p_{3/2}}$ & &     4.5  & &      3.1      & &   6.18\\
  & $\Delta E_{d_{3/2}-d_{5/2}}$ & &     3.9  & &      3.6      & &   5.08\\
~\\
p.-h. gaps: &
    $\Delta E_{s_{1/2}-p_{1/2}}$ & &    14.6  & &     12.2      & &  12.4\\
  & $\Delta E_{d_{3/2}-p_{1/2}}$ & &    19.3  & &     16.5      & &  16.6\\
  & $\Delta E_{d_{5/2}-p_{1/2}}$ & &    15.4  & &     13.0      & &  11.5\\
 \hline \hline
\end{tabular}
\end{center}
\caption[]{Spin-orbit splittings and ph gaps of ${}^{16}$O (in MeV) obtained for the $V_{\text{UCOM}}$
  and Argonne {\em v18} interactions~\cite{bar.06}.
 A G-matrix resummation of ladders outside the model space is employed for {\em v18}.
 The experimental values refer to the spectra of the corresponding states in ${}^{17}$O and ${}^{15}$O.}
\label{sosplit}
\end{table}

\begin{figure}[t]
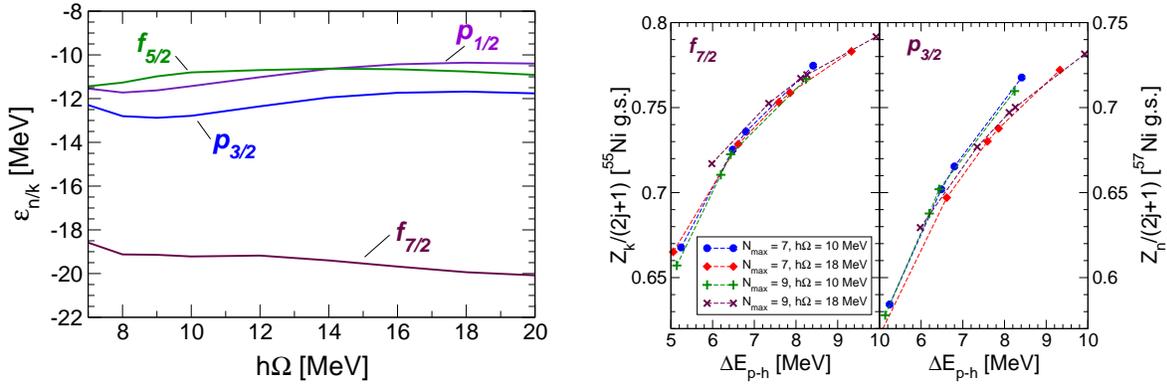

\begin{center}
\includegraphics[width=0.45\columnwidth,clip=true]{Ni56_spe.eps}
  \hspace{.5cm}
\includegraphics[width=0.47\columnwidth,clip=true]{Ni56_Kdep_f7p3.eps}
\end{center}
\caption{Left: neutron quasiparticle energies for $^{56}$Ni obtained for the N3LO interaction as
       a function the the oscillator frequency.
      A monopole correction was added to constrain the ph gap to its experimental value
      at $\hbar\Omega$=10~MeV.
       Right: correlation between the spectroscopic factors for the $1p_{3/2}$ and $0f_{7/2}$ orbits
       and the ph gap. Each line correspond to different $\hbar\Omega$ and sizes of the model space and
       is obtained by varying the monopole correction term.
  }
\label{fig:Ni56spe}
\end{figure}

Recent improvements in the numerical codes for calculating the FRPA self-energy have allowed to extend the above calculations to larger nuclei in the $0p1f$ shell. For $^{56}$Ni, calculations were performed in spaces up to ten oscillator shells~\cite{bar.09c}. In this work, the chiral N3LO interaction~\cite{ent.03} was modified by a simple correction of the monopole interaction. This correction is needed with most realistic two-nucleon Hamiltonians in order to reproduce correctly the experimental gap between the $1p_{3/2}$ and the $0f_{7/2}$ orbits~\cite{ANT.05}.
 The dependence of single particle energies on the oscillator frequency, $\hbar\Omega$, is given by the left panel of Fig.~\ref{fig:Ni56spe}. The results show limited deviations for a large range of frequencies. Part of the residual dependence is also due to the prescription used for the monopole correction, which is also dependent on $\hbar\Omega$.
 Interestingly enough, it is found that the calculated spectroscopic factors for these single particle states are only sensible to the ph energy gap. This is shown by the right panel of Fig.~\ref{fig:Ni56spe}, where calculation are reported for different sizes of the model spaces and monopole corrections~\cite{bar.09c}.
It follows that, even if calculations are not perfectly converged, stable predictions for the spectroscopic factors can be obtained by constraining the ph gap.
A study of the contributions to the spectroscopic factors from different types of correlations is reported in the next section.

\section{Results for Spectroscopic Factors}
\label{sec:sf}

Spectroscopic factors for particle and hole states around closed shell
nuclei are of particular interest since they carry
important information regarding nuclear structure~\cite{dic.04}.
Strong deviations from unity signal the onset of substantial correlation
effects and imply the existence of non trivial many-body dynamics.
Understanding how spectroscopic factors change when moving close to the drip lines is important
to constrain theoretical models of radioactive isotopes.

\subsection{Role of Long-Range Correlations}
\label{sec:sfNi56}

\begin{table}[t]              
\begin{tabular}{rlcccccccc}
\hline                        
\hline                        
                  & &  FRPA     &    Exp.   &~&  FRPA     &    FRPA   & $\Delta Z_\alpha$ & FRPA &    \\
                  & &           &           &~&   (SRC)   &           &                   & +$\Delta Z_\alpha$  &  Exp. \\
                  &&(10 shells)&            & &(10 sh.) &(10 sh.)&   (pf shell)      &  (10 sh.)      & \\
\hline                        
$^{57}$Ni:
& $\nu 1p_{1/2}$    &  -11.43   &  -9.134   & &   0.96    & 0.63      & -0.02             & 0.61     & \\
& $\nu 0f_{5/2}$    &  -10.80   &  -9.478   & &   0.95    & 0.59      & -0.04             & 0.55     & \\
& $\nu 1p_{3/2}$    &  -12.78   &  -10.247  & &   0.95    & 0.65      & -0.03             & 0.62    &   0.58(11) \\
$^{55}$Ni:
& $\nu 0f_{7/2}$    &  -19.22   &  -16.641  & &   0.95    & 0.72      & -0.03             & 0.69     &  \\
\\
$^{57}$Cu:
& $\pi 1p_{1/2}$    &   -1.28   &  +0.417   & &   0.96    & 0.66      & -0.04             & 0.62     & \\
& $\pi 0f_{5/2}$    &   -0.58   &           & &   0.96    & 0.60      & -0.02             & 0.58     &  \\
& $\pi 1p_{3/2}$    &   -2.54   &  -0.695   & &   0.96    & 0.67      & -0.02             & 0.65     &  \\
$^{55}$Co:
& $\pi 0f_{7/2}$    &   -9.08   &  -7.165   & &   0.95    & 0.73      & -0.02             & 0.71     &  \\
\hline                        
\hline                        
\end{tabular}
 \caption[]{Energies (in MeV) and spectroscopic factors (as a fraction of
    the independent-particle model) for transitions to the $1p0f$ valence orbits of $^{56}$Ni, obtained
    for $\hbar\Omega=10$~MeV~\cite{bar.09c,bar.09.he4}.
     The third and fourth columns correspond to the sole contribution of correlations from outside the
     model space (SRC) and to the full FRPA result.
    The corrections $\Delta Z_\alpha$ are obtained by comparing shell model and FRPA in the $pf$ shell alone
    and are added in the sixth column.
    The experimental single-particle energies are taken from~\cite{toi.96}.
    The measured spectroscopic factor for transfer between the ground states
    of $^{57}$Ni and $^{56}$Ni is from Ref.~\cite{yur.06}.
    }
\label{tab:sf}
\end{table}

Spectroscopic factors for the quasiparticle and quasihole peaks around $^{56}$Ni are reported
in Tab.~\ref{tab:sf}~\cite{bar.09.he4,bar.09c}. These correspond to the single particle levels
displayed in Fig.~\ref{fig:Ni56spe} for $\hbar\Omega$=10~MeV and were calculated for the
same monopole corrected N3LO interaction (which reproduces the ph gap at Z,N=28).
Theoretically, spectroscopic factors are defined as the norm of the overlap function $\psi_\alpha({\bf r})$
for the nucleon transfer to a given quasiparticle state $\alpha$. This can be obtained from the nuclear
self-energy by the relation~\cite{dic.04,dicvan}
\begin{equation}
Z_\alpha  = \int d^3 r \; \vert \psi_\alpha ({\bf r}) \vert^2 =
  \left.
  \frac{1}{  1   -
  \frac{\partial \Sigma^\star_{\hat\alpha \hat\alpha}(\omega)}{\partial \omega}
  } \right\vert_{\omega=\varepsilon_\alpha}
\label{sf}
\end{equation}
where $\Sigma^\star_{\hat\alpha \hat\alpha}(\omega) \equiv <\hat{\psi}_\alpha|\Sigma(\omega)|\hat{\psi}_\alpha>$ is the expectation value of the
self-energy calculated for the overlap function itself but normalized to unity ($\int|\hat{\psi}_\alpha|^2=1$).

The third column in Tab.~\ref{tab:sf} shows the results obtained from the self-consistent self-energy calculated by
including in Eq.~(\ref{sf}) only the contributions induced by degrees of freedom from outside the model space.
As the N3LO interaction is rather soft compared to other realistic models the effects of short-range correlations  (SRC) are small.
For interactions such as the Bonn or Argonne models, one should expect a quenching up to about 10\% or more~\cite{dic.04,rio.09}.
These early results are confirmed by recent electron scattering experiments~\cite{roh.04,bar.04,SciSRC}.
The complete FRPA results, in the model space of 10 oscillator shells plus SRC, are given in the
fourth column. For the transition between the $^{56}$Ni and $^{57}$Ni ground states our result
agrees, within the error bar, with the experimental value of Ref.~\cite{yur.06}.
The remaining orbits have similar quenching and  are in line with values expected
for stable closed-shell nuclei.
Therefore, for this nucleus the dominant mechanisms in reducing the spectroscopic strength
is to be looked for in long-range correlations, involving configurations among a several
shells around the Fermi surface.

We now turn to the question of what are the effects of correlations induced by
configurations beyond 2p1h and 2h1p. Since  extra ph excitations are induced by the
self-consistency approach one can expect the SC-FRPA to be accurate
for closed shell nuclei, even though one does not perform
a full configuration mixing.
To show that this is the case, we performed both SC-FPRA and shell model calculations
in the $1p0f$ shell~\cite{bar.09d}. This was done using the same interaction and single particle orbits
employed in the large oscillator basis. The ANTOINE shell model code was used~\cite{ANT.99,ANT.05}.
As an example, the spectroscopic factor for the $p_{3/2}$ ground state of $^{57}$Ni, was predicted
to be 0.82 by FRPA and 0.79 by the shell model with up to 6p6h states.
Comparing with the full space result of 0.65 in Tab.~\ref{tab:sf}, one can infer that
about half of the quenching is driven by degrees of freedom from outside the $pf$ space.
The difference between shell model and FRPA, $\Delta Z_\alpha$=0.79-0.82=-0.03, is rather small as expected.
The fifth and sixth columns in the table show the corrections $\Delta Z_\alpha$ for all single particle orbits and the resulting
spectroscopic factors when the $\Delta Z_\alpha$s are subtracted from the full space FRPA results.

It remains clear that the extra correlations from the shell model (i.e., not included by FRPA) will become crucial in open shell nuclei.
Nevertheless, these results give confidence that calculations based on Green's functions theory can lead to highly accurate
results for quasiparticles in closed shells.
This opens a new way to derive the properties of shell model Hamiltonians from realistic forces.

\subsection{Proton-Neutron Asymmetry}

Understanding how spectroscopic factors change when moving close to the drip lines is important
to  constrain theoretical models of radioactive isotopes.
First information on these features has recently become available
using one-nucleon knockout experiments in inverse kinematics.
In general, it is found that spectroscopic factors do change with proton-neutron asymmetry
and the quenching of quasiparticle orbits close to the Fermi surface become
stronger with increasing separation energy~\cite{gad.08}.

\begin{figure}[t]
\begin{center}
\includegraphics[width=0.56\columnwidth,clip=true]{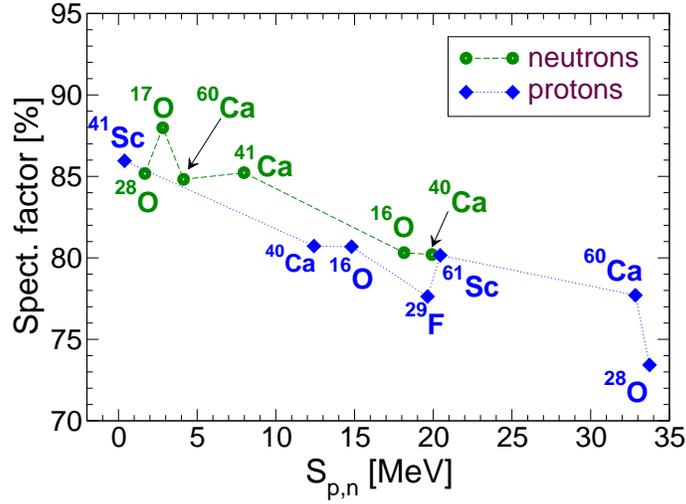}
\end{center}
\caption{Spectroscopic factors obtained from partially self-consistent FRPA. All numbers are
  given as a fraction of the independent particle model value and refer to transitions from ground
  state to ground state. The points refer to knockout of a nucleon from the
  isotope indicated nearby. The lines are a guide to the eye.}
\label{SFasymm}
\end{figure}

Figure~\ref{SFasymm} shows FRPA results for the spectroscopic factors
of quasiparticles around $^{\rm 16,28}$O and $^{\rm 40,60}$Ca~\cite{bar.09b}.
These are based on the realistic chiral N3LO interaction and employ also a G-matrix
to account for the effects of short-range correlations.
A dependence on proton-neutron asymmetry is indeed observed in the FRPA,
with the spectroscopic factors becoming smaller with increasing nucleon
separation energy.  A dispersive optical model analysis, which is constrained
to data up to $^{\rm 48}$Ca, has also been extrapolated to neutron rich Ca
isotopes with similar findings~\cite{cha.06}.
 However, for both analyses the change in magnitude is significantly
smaller than the one deduced from direct knockout data~\cite{gad.08}.

We stress that the calculations of Fig.~\ref{SFasymm} fail in reproducing
energy gaps across shells and the corresponding location of giant resonances. This is a typical
finding for soft low-energy interactions (see, for example, $V_{\text{UCOM}}$ in Tab.~\ref{sosplit})
which are incomplete without multi-nucleon forces.
Based on the findings of Sec.~\ref{sec:sfNi56}, collective excitations from Fig.~\ref{frpa_diags}c
are the most important degrees of freedom governing the reduction of spectroscopic factors~\cite{dic.04,bar.09b,bar.09.he4}.
These are properly accounted for by the FRPA approach.
However, \emph{either} realistic three-nucleon forces \emph{or} phenomenological corrections will be needed
to achieve reliable predictions.
It is plausible that the dependence on asymmetry seen in Fig.~\ref{SFasymm}
may become more substantial once FRPA calculations with improved interactions will be
available.

\section{Conclusions}

We reported about ongoing calculations of the binding energy of $^4$He using the FRPA method within the framework of self-consistent Green's functions theory.  The results show that this approach can achieve high accuracy and provides a promising method for \emph{ab-initio} calculations in medium-heavy nuclei.
 As an example of its application, we discussed quasiparticle states in $^{16,28}$O, $^{40,60}$Ca and $^{56}$Ni and the N/Z dependence of the spectral strength.
Spectroscopic factors of nuclei around $^{56}$Ni are found to receive very small corrections from missing many-body correlations.
This approach opens a new way to derive the properties of shell model Hamiltonians from realistic nuclear forces.

\section*{Acknowledgments}
It is a pleasure to thank W.~H.~Dickhoff, M.~Hjorth-Jensen, K. Langanke, G.~Mart\'{\i}nez-Pinedo, T.~Otsuka and D.~Van Neck for several stimulating discussions.
The nuclear matrix elements employed in all calculations were obtained using the CENS code~\cite{CENS1,CENS2}.
This work was supported by the Japanese Ministry of Education, Science and Technology (MEXT) under KAKENHI grant no. 21740213.


\begin{thebibliography}{99}
\bibitem{sch.02}  J. P. Schiffer \emph{ et al.}, Phys. Rev. Lett. {\bf 92}, 162501 (2004).
\bibitem{taka.05} T.~Otsuka, Phys. Rev. Lett. {\bf 95}, 232502 (2005).
\bibitem{gad.08}  A.~Gade,  et al., Phys.\ Rev.\ C {\bf 77}, 044306 (2008).
\bibitem{he_bench}  H. Kamada \emph{ et al.}, Phys. Rev. C {\bf 64}, 044001 (2001)
\bibitem{hag.07}  G. Hagen, D. J. Dean, M. Hjorth-Jensen, T. Papenbrock, and A. Schwenk, Phys. Rev. C {\bf 76}, 044305 (2007).
\bibitem{hag.08}  G. Hagen, T. Papenbrock, D. J. Dean, and M. Hjorth-Jensen, Phys. Rev. Lett. {\bf 101}, 092502 (2008).
\bibitem{hag.09}  G. Hagen, T. Papenbrock, D. J. Dean, M. Hjorth-Jensen, and B. Velamur Asokan, Phys. Rev. C {\bf 80}, 021306(R) (2009).
\bibitem{bar.06}  C.~Barbieri, Phys.~Lett.~{\bf B643}, 268 (2006).
\bibitem{bar.09c} C.~Barbieri and M.~Hjorth-Jensen, Phys. Rev. C{\bf 79}, 064313 (2009).
\bibitem{dicvan}  W.~H.~Dickhoff and D.~Van~Neck, \emph{ Many-Body Theory Exposed!} (World Scientific, Singapore, 2005).
\bibitem{bar.01}  C. Barbieri and W. H. Dickhoff, Phys. Rev. C {\bf 63}, 034313 (2001).
\bibitem{bar.07}  C.~Barbieri, D. Van Neck and W.~H.~Dickhoff, Phys.\ Rev.\ A{\bf 76}, 052503 (2007).
\bibitem{Bog.03}  S.K. Bogner, T.T.S. Kuo, A. Schwenk, Phys. Rept. {\bf 386}, 1 (2003).
\bibitem{Bog.07}  S. K. Bogner et al., Nucl. Phys. {\bf A 784}, 79 (2007).
\bibitem{bof.71}  S.~Boffi, Nuovo Cimento Lettere {\bf 1}, 931 (1971).
\bibitem{bar.09.he4} C.~Barbieri, in preparation.
\bibitem{tro.05}  A.~B.~Trofimov and J.~Schirmer, J. Chem. Phys. {\bf 123}, 144115 (2005).
\bibitem{bar.09a} C. Barbieri and D. Van Neck, AIP Conf. Proc. {\bf 1120} 104 (2009); and in preparation.  
\bibitem{bay.61}  G.~Baym and L.~P.~Kadanoff, Phys.~Rev.~{\bf 124}, 287 (1961).
\bibitem{bay.62}  G.~Baym, Phys.~Rev.~{\bf 127}, 1391 (1962).
\bibitem{rot.05}  R. Roth, H. Hergert, P. Papakonstantinou, T. Neff, and H. Feldmeier, Phys. Rev. C {\bf 72}, 034002 (2005)
\bibitem{wir.95}  R.~B.~Wiringa, V.~G.~J.~Stoks, and R.~Schiavilla, Phys.\ Rev.\ C {\bf 51}, 38 (1995).
\bibitem{geu.96}  W.~J.~W.~Geurts, K.~Allaart, W.~H.~Dickhoff, and H.~M\"uther,  Phys. Rev. C {\bf 53}, 2207 (1996).
\bibitem{pie.93}  S.~C.~Pieper and V.~R.~Pandharipande, Phys. Rev. Lett. {\bf 70}, 2541 (1993).
\bibitem{rot.06}  R. Roth, P. Papakonstantinou, N. Paar, H. Hergert, T. Neff, and H. Feldmeier, Phys. Rev. C{\bf 73}, 044312 (2006)
\bibitem{rot.09}  R. Roth, J. R. Gour, and P. Piecuch, Phys. Rev. C {\bf 79}, 054325 (2009)
\bibitem{ent.03}  D. R. Entem and R. Machleidt, Phys. Rev. C {\bf 68}, 041001(R) (2003).
\bibitem{ANT.05}  E. Caurier \emph{ et al.}, Rev. Mod. Phy. {\bf 77}, 427 (2005).
\bibitem{dic.04}  W. H. Dickhoff and C. Barbieri, Prog.~Part.~Nucl.~Phys.~{\bf 52}, 377 (2004).
\bibitem{toi.96} R.~B.~Firestone, V.~S.~Shirley, C.~M.~Baglin, S.~Y.~Frank~Chu, and J.~Zipkin, Table of Isotopes, 8th ed. (Wiley Interscience, New York, 1996).
\bibitem{rio.09}  A.~Rios, A.~Polls, and W. H. Dickhoff, Phys. Rev. C {\bf 79}, 064308 (2009).
\bibitem{roh.04}  D. Rohe \emph{ et al.}, Phys. Rev. Lett. {\bf 93}, 182501 (2004).
\bibitem{bar.04}  C. Barbieri and L.~Lapik\'{a}s, Phys. Rev. C {\bf 70}, 054612 (2004).
\bibitem{SciSRC}  R. Subedi \emph{ et al.}, Science {\bf 320}, 1476 (2008).
\bibitem{yur.06}  K.~L.~Yurkewicz \emph{ et al.}, Phys. Rev. C {\bf 74}, 024304 (2006).
\bibitem{bar.09d} C.~Barbieri, arXiv:0909.3040 [nucl-th].
\bibitem{ANT.99}  E. Caurier and F. Nowacki, Acta Phys. Polonica 30, 705 (1999).
\bibitem{bar.09b} C.~Barbieri and W.~H.~Dickhoff, Int. J. Mod. Phys. A24, 2060, (2009).
\bibitem{cha.06}  R.~J.~Charity  \emph{ et al.}, Phys.\ Rev.\ Lett. {\bf 97}, 162503 (2006); Phys.\ Rev.\ C {\bf 76}, 044314 (2007).
\bibitem{CENS1}   M.~Hjorth-Jensen \emph{ et al.}, Phys. Rep. {\bf 261} 125 (1995).
\bibitem{CENS2}   T. Engeland, M.Hjorth-Jensen and G.R. Jansen, CENS, a Computational Environment for Nuclear Structure. 



\end{thebibliography}
\end{document}